# A hidden low-temperature transformation pathway in compositionally complex materials

**Yujiao Li[1*], Elaheh Akbarnejad[2], Quentin Bizot[3], Aleksander Kostka[1], Natalia Pukhareva[2], Ridha Zerdoumi[2], Alan Savan[2], Matous Mrovec[3], Ralf Drautz[3], Baptiste Gault[4,5], Dierk Raabe[4], Alfred Ludwig[1,2,6*]**

Corresponding authors (*): yujiao.li@rub.de; alfred.ludwig@rub.de

[1] Center for Interface-Dominated High-Performance Materials (ZGH), Ruhr University Bochum, Bochum, Germany
[2] Chair for Materials Discovery and Interfaces, Institute for Materials, Ruhr University Bochum, Bochum, Germany
[3] Interdisciplinary Centre for Advanced Materials Simulation (ICAMS), Ruhr University Bochum, Bochum, Germany
[4] Max Planck Institute for Sustainable Materials, Düsseldorf, Germany
[5] Groupe de Physique des Matériaux, UMR 6634, Université de Rouen Normandie, CNRS, INSA Rouen Normandie, Rouen, France
[6] Research Center Future Energy Materials and Systems (RC FEMS), Ruhr University Bochum, Bochum, Germany

**Abstract**

Most compositionally complex materials (CCMs, frequently referred to as high entropy alloys) are metastable[1–3] and their attractive properties[4–8] often belong to kinetically trapped states. However, pathways towards lower-free-energy phase states governing long-term stability, can remain hidden because diffusion-controlled atomic redistribution is too slow to be revealed at experimentally accessible timescales[9–11]. This blind spot is acute in CCM design: enormous compositional spaces[12] are screened for performance, yet the low-temperature kinetics and the associated transformation pathways determining whether that performance persists are rarely considered in material selection. Here we use defect-rich nanoscale volumes coupled with atom-probe tomography[13] to access and reconstruct the hidden phase-evolution pathway in a metastable $Ag_{24}Au_{20}Pd_{50}Pt_6$ electrocatalyst, without relying on elevated temperatures to accelerate the transformation. By varying microstructural starting state, annealing temperature and time, we reveal precipitation of a Pt-rich phase within the fcc matrix, its coarsening and re-homogenization. The Pt-rich phase recurs after homogenization with delayed kinetic accessibility, while prolonged annealing extends the pathway to 300°C. Atomistic simulations independently predict the same Pt-rich phase selection. The transformation is accompanied by a 3.7-fold loss of catalytic activity for hydrogen evolution. These results establish hidden phase-evolution pathways as a materials-design variable: resolving them can guide the selection of metastable CCMs not only for their as-synthesized properties, but also for the phase states and associated functionalities they may access over time.

## Introduction

Compositionally complex materials (CCMs) access large compositional spaces in which mechanical[4–7,14,15] and functional properties[8,12,16,17] can be tuned beyond conventional alloy design. However, most of them are synthesized in metastable states[1–3,18,19]. Their long-term stability and thus useful lifetime are governed not only by their as-synthesized state, but by the pathway through which that state evolves towards lower-free-energy phase states. At either kinetic extreme, the consequences of metastability are comparatively clear: quenched martensitic steels evolve readily during low-temperature tempering[20], whereas kinetically frozen materials such as diamond persist for technologically relevant timescales despite their thermodynamic metastability[21]. The challenge is most acute for metastable CCMs, which occupy an intermediate kinetic regime: transformations are too slow to be observed directly in routine experiments, yet not necessarily too slow to alter properties during service. This timescale mismatch leaves their relevant phase-evolution pathways experimentally hidden (Fig. 1a). The central challenge is therefore not simply to identify an equilibrium state, but to reveal the equilibrium-seeking pathways accessible from a specific microstructural history.

Experimental access to such pathways is limited by diffusion-controlled redistribution in concentrated solid solutions[10]. The extensively studied Cantor alloy[9], a prototypical CCM, provides a striking example: phase decomposition has been observed only after approximately 500 days of annealing at intermediate homologous temperatures ($T_{\mathrm{hom}}=T/T_{\mathrm{M}}\approx0.45$-$0.57$, where $T_{\mathrm{M}}$ is the melting point)[11]. At lower temperatures, where the thermodynamic driving force for decomposition may increase, progressively slower atomic mobility makes experimental access increasingly difficult. This divergence between thermodynamic driving force and kinetic accessibility leaves low-temperature phase-evolution pathways unresolved outside the conventional experimental window (Fig. 1a). Raising the annealing temperature accelerates transport[18,22,23], but can also shift the alloy into a different phase-stability regime[24]. High-temperature behavior therefore cannot be used as a simple proxy for low temperature phase evolution. Atomistic simulations can predict thermodynamic tendencies[25], but they do not by themselves establish the route taken by a real material with its processing-inherited microstructure.

A suitable experiment must therefore relax kinetic constraints sufficiently to make otherwise inaccessible phase evolution observable, while preserving the thermodynamic driving forces governing that evolution. Our earlier works[26,27] on the nanocrystalline thin film CCMs established the acceleration principle: defect-rich nanoscale volumes containing abundant grain boundaries (GBs) and vacancies[28,29] can make otherwise slow decomposition observable at low temperatures. However, those experiments were limited by alloy-substrate interactions at 300°C[30] and did not systematically reconstruct an evolution pathway across a broad temperature range. To overcome these limitations, we thermally oxidize the Si microtip arrays before alloy deposition, creating a $SiO_2$ diffusion barrier that suppresses interactions up to 700°C (Methods), enabling characteristic metallurgical evolution across an extended temperature range to be captured within a similar experimental platform, which we term here *nanolab metallurgy* (Extended Data Fig. 1 and 2). Each *nanolab* is a coated $SiO_2$/Si microtip array comprising 36 atom probe tomography (APT)-ready specimens (~0.2 µm$^3$ each). After each heat-treatment step, a subset of the specimens is consumed by APT, a technique that provides compositional mapping in 3D with sub-nanometer resolution[31–35] (Methods). The remaining specimens undergo subsequent annealing and hence retain the accumulated thermal history (Fig. 1b). Assuming that the initial microstructure of all specimens is equivalent, this serial APT sampling reveals successive states arising over time at higher temperature, enabling reconstruction of a phase-evolution pathway rather than merely observing accelerated isolated transformations.

## Results and Discussion

Here we apply this strategy to metastable $Ag_{24}Au_{20}Pd_{50}Pt_6$, an alloy selected from a combinatorial Ag-Au-Pd-Pt thin-film materials library designed to explore how composition influences the electrocatalytic activity of polyelemental surfaces (Extended Data Fig. 3). This specific composition was chosen as a model catalytic system near the competition between solid-solution retention and phase separation (Extended Data Fig. 4). CCMs have recently proven to be promising catalysts, for different key electrochemical reactions including the hydrogen-evolution reaction (HER)[36]. However, annealing-induced phase evolution in metastable $Ag_{24}Au_{20}Pd_{50}Pt_6$ is accompanied by an approximately 3.7-fold loss of HER activity (Extended Data Fig. 3), highlighting the relevance of the hidden transformation to functional persistence. Microstructural starting state as well as annealing temperature and time provide three complementary experimental probes (Fig. 1c; Extended Data Fig. 5) that allow us to reconstruct the phase-evolution pathway and reveal the underlying thermodynamic tendency and the history-dependent kinetic accessibility that governs when the transformation becomes experimentally observable (Fig. 1c). All routes lead to the same Pt-rich phase-selection tendency despite differences in temperature, time and microstructural history, down to $T_{hom} \approx 0.36$. Machine-learning-enabled atomistic simulations independently predict a Pt-rich phase as the low-temperature thermodynamic equilibrium state, consistent with the recurrent experimental phase-selection. Together, these results show that the long-term evolution of metastable CCMs reflects thermodynamic phase selection constrained by history-dependent kinetic accessibility. Access to these pathways is therefore critical for selecting and designing CCMs whose attractive properties persist under service conditions.

**Fig. 1 Reconstructing a hidden low-temperature transformation pathway in a metastable CCM. a,** Schematic time-temperature representation of phase evolution in a metastable bulk CCM solid solution. High-temperature phase evolution is accessible within a limited experimental window but may enter a different phase-stability regime, whereas low-temperature phase evolution remains kinetically inaccessible and its outcome experimentally hidden. **b**, Nanolab comprising a $SiO_2$/Si microtip array with 36 defect-rich nanoscale alloy volumes (~0.2 µm$^3$ each) with abundant grain boundaries and vacancies that serve as individual atom probe tomography (APT) specimens. Sequential heat treatment and destructive APT analysis of selected specimens reconstruct successive states ($S_0$, $S_1$,…), while the remaining specimens retain the accumulated thermal history for subsequent annealing; grey tips denote specimens consumed by APT analysis. **c**, Three complementary probes - annealing temperature, microstructural starting state and time - reconstruct the phase-evolution pathway in $Ag_{24}Au_{20}Pd_{50}Pt_6$ (Extended Data Fig. 5). The experiment reveals recurrent Pt-rich phase selection, accompanied by an approximately 3.7-fold loss in hydrogen-evolution activity (Extended data Fig. 3). Atomistic simulations independently predict the same low-temperature phase-selection tendency.

We first used temperature as a probe to progressively reveal the phase evolution concealed within the as-deposited $Ag_{24}Au_{20}Pd_{50}Pt_6$ alloy (Fig. 2a). The as-deposited state, characterized by transmission electron microscopy (TEM; see Methods), is a nanocrystalline single-phase face-centered cubic (fcc) solid solution with a grain size of approximately 10 – 20 nm (Fig. 2b and Extended Data Fig. 2), exhibiting nanoscale chemical modulation without a detectable secondary phase. Such modulations in sputter-deposited noble-metal alloys have been associated with deposition rate and element-dependent surface mobility[37,38]. High-resolution TEM (HRTEM) resolves two differently oriented fcc grains, while continuous lattice fringes across the chemical modulations show that the compositional fluctuations do not disrupt the fcc lattice continuity (Fig. 2b). After annealing at 200°C for 1 h, the modulation persists without formation of a second phase (Fig. 2c).

Increasing temperature progressively reveals the evolution towards phase separation. At 400°C, Pt-rich nuclei first emerge preferentially at GBs, accompanied by pronounced Ag depletion (Fig. 2d). At 500°C, their average size increases by approximately fivefold while their number density remains comparable, consistent with the growth of existing nuclei rather than extensive new nucleation (Fig. 2e). Annealing at 600°C produces pronounced coarsening and continued Pt enrichment[39,40], whereas Ag depletion for the Pt-rich phase is already largely established at 500°C (Figs. 2f,g).

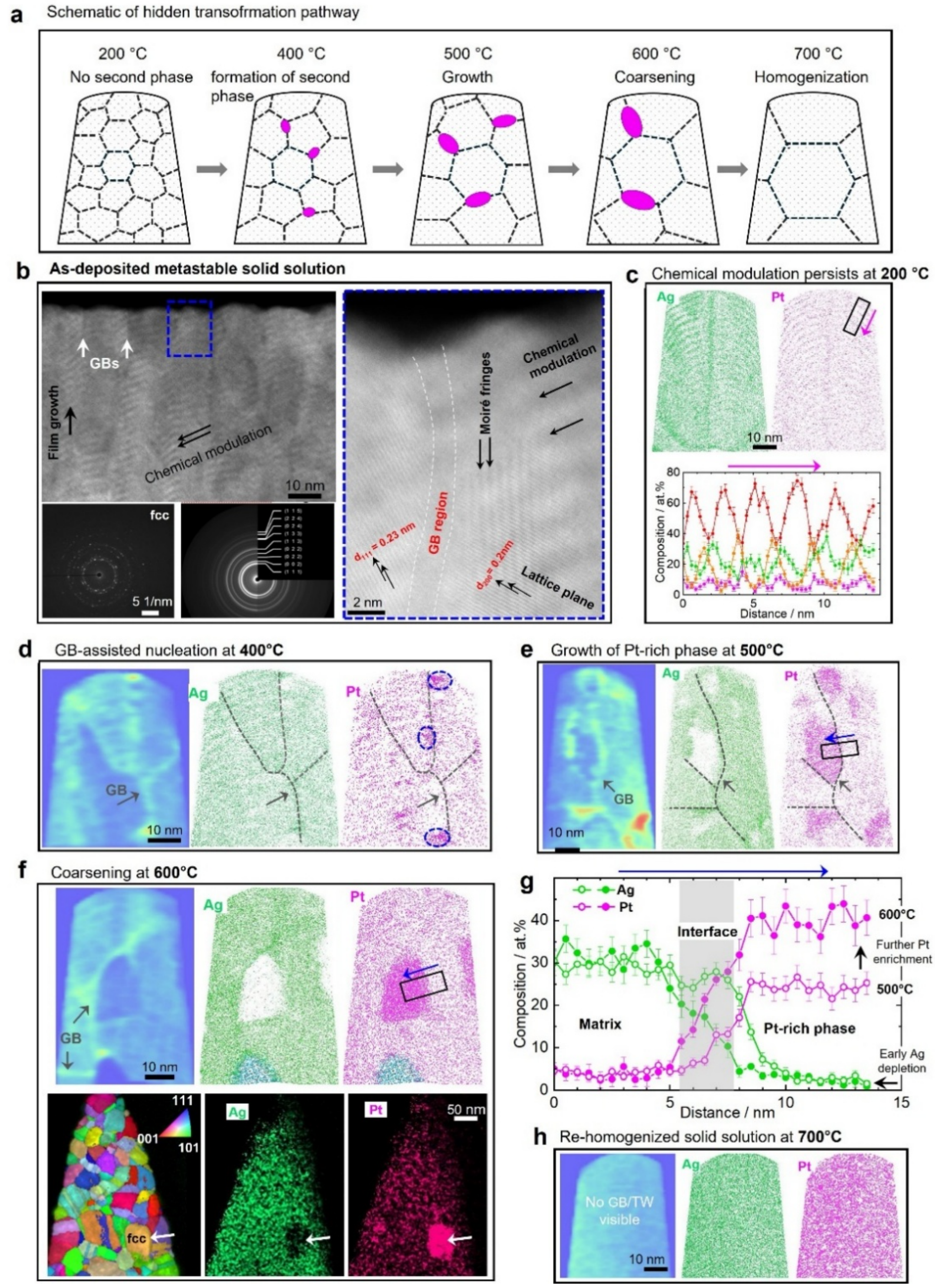

**Fig. 2 Temperature progressively reveals a hidden decomposition-homogenization pathway in metastable $Ag_{24}Au_{20}Pd_{50}Pt_6$ during stepwise annealing from 100 to 700°C.** Each temperature was maintained for 1 h. **a,** Schematic of the phase-evolution pathway reconstructed across the temperature series. **b**, As-deposited fcc solid solution. TEM and HRTEM images reveal a nanocrystalline microstructure containing GBs and nanoscale chemical modulation as well as three characteristic length scales: lattice fringes corresponding to fcc {111} (~0.23 nm) and {200} (~0.20 nm) planes, superimposed Moiré fringes arising from overlapping nanocrystals (~0.5 nm), and longer-range chemical modulation (~3 nm). **c,** APT results after annealing at 200°C with chemical modulation persisted. **d,** Pt-rich nuclei first appear at GBs at 400°C, accompanied by Ag depletion. Local-magnification contrast[39,40] in the 2D atomic-density map reveals GBs. **e,** Growth of the Pt-rich phase at 500°C along GBs. The highlighted region indicates the location used for quantitative composition profiling. **f,** Pt-rich phase coarsens at 600°C and is confirmed by APT and TEM as a chemically distinct fcc-derived crystal structure. **g,** Composition profiles across the matrix-Pt-rich phase interface at 500°C (open circles) and 600°C (solid circles). **h,** Dissolution of the Pt-rich phase and re-homogenization at 700°C. Only Ag and Pt atom maps are shown, as they capture the dominant compositional redistribution during phase evolution. Au and Pd maps, corresponding composition profiles and the 2D density color scale are provided in Extended Data Figs. 6 and 7.

Semi-correlative APT and TEM[41] after annealing at 600°C confirm that the chemically distinct Pt-rich phase is crystalline and fcc-like. At 700°C, the Pt-rich phase redissolves, producing a chemically more homogeneous (compared to the as-deposited state) single-phase CCM accompanied by substantial grain growth, with no GBs observed within the APT field of view (Fig. 2h). The temperature sequence reveals a decomposition-homogenization pathway from the metastable as-deposited state. However, this raises the question of whether Pt-rich phase formation is intrinsic to the alloy or inherited from its as-deposited starting state.

To determine whether Pt-rich phase formation depends on the as-deposited starting state, we used the 700°C-homogenized state (Fig. 2h, $S_7$ in Fig. 1c) as a second starting state (Fig. 3). This state is a chemically homogeneous single-phase solid solution with substantially enlarged grains and no GBs observed within the analyzed APT volume. Composition profiles are uniform, and the experimental binomial frequency distributions closely overlap those expected for a random solid solution (Fig. 3a). Subsequent 1-h annealing of this homogenized state at 400°C and 500°C produced the states denoted H400 and H500, respectively. At H400, no Pt-rich phase is detected (Fig. 3b), in contrast to its formation after 1 h at 400°C from the nanocrystalline as-deposited state (Fig. 2d). At H500, however, the Pt-rich phase re-emerges at GBs (Fig. 3c), showing that homogenization shifts its detectable formation to higher temperatures within the 1-h annealing window. This shift is consistent with reduced kinetic accessibility after homogenization, which induces substantial grain coarsening and alters the initial chemical heterogeneity, although their individual contributions cannot be distinguished. Si segregation at GBs in H400 and H500 (Extended Data Fig. 9), possibly introduced during 700°C homogenization, cannot account for Pt-rich phase formation, which occurs at GBs at 400 and 500°C before homogenization without detectable Si segregation.

Chemical heterogeneity was quantified using the Pearson coefficient $\mu$ derived from binomial frequency-distribution analysis[42] (Fig. 3 and Methods), with lower $\mu$ indicating elemental distributions closer to random. The element-specific $\mu$ values increase with the formation of Pt-rich phase, decrease upon its dissolution and homogenization at 700°C, remain low at H400 and increase again with the re-emergence of the Pt-rich phase at H500. The mean $\mu$ across Ag, Au, Pd and Pt decreases from 0.49 in the as-deposited state to 0.14 after homogenization, quantitatively distinguishing the two starting states.

Despite these differences, both starting states repeatedly evolve towards a Pt-rich phase, whereas its observable onset within the 1-h annealing window shifts from 400°C to 500°C after homogenization. The starting state therefore changes the conditions under which the pathway becomes experimentally accessible, rather than the observed direction of phase evolution. If this observable onset is kinetically defined, can longer annealing reveal the same phase-evolution tendency at lower temperature?

We therefore used a second nanolab to probe extended annealing at 300°C (Extended Data Fig. 5b). As established from the first nanolab, no Pt-rich phase was detected after 1 h in this case (Extended Data Fig. 8), consistent with the low chemical heterogeneity quantified by $\mu$ (Fig. 3d), whereas extended annealing for 68 h produced a Pt-rich phase with characteristic sizes comparable to those formed after 1 h at 400°C (Fig. 4a). Prolonged annealing thus reveals a lower-temperature continuation of the Pt-rich phase-selection pathway that remains hidden within the 1 h window. This time-temperature equivalence identifies the apparent onset temperature as a kinetic observation boundary rather than, by itself, a thermodynamic phase boundary.

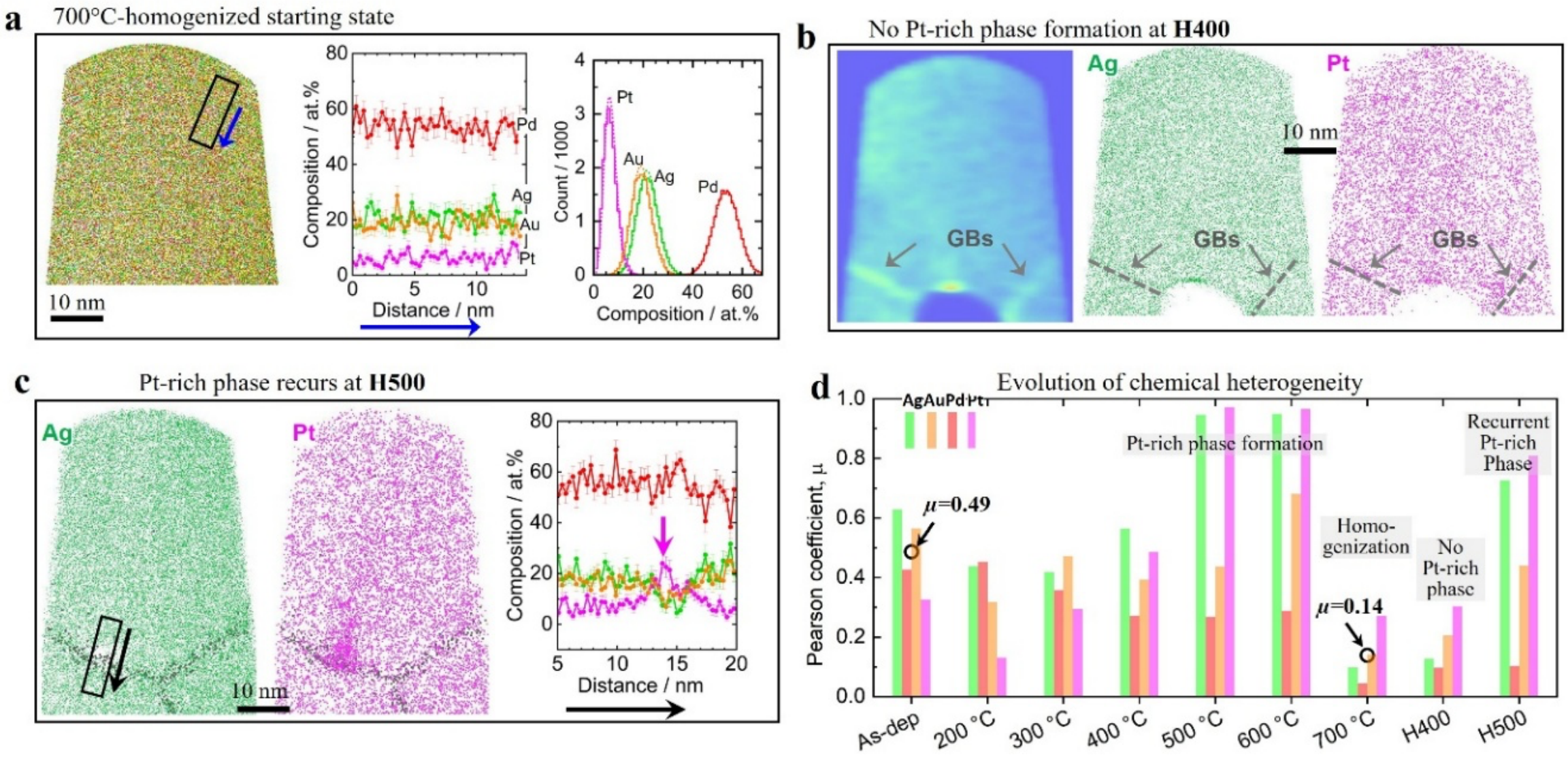

**Fig. 3 Starting state controls the kinetic accessibility of phase separation**. **a**, Homogenized starting state obtained after annealing at 700°C for 1 h. APT reconstruction, composition profile and binomial frequency analysis show a chemically homogeneous solid solution with near-random elemental distributions (solid, and experimental frequency; dashed, theoretical binomial distributions). H400 and H500 denote reheating of this homogenized state at 400°C and 500°C for 1 h, respectively. **b**, No detectable Pt-rich phase is observed at H400. **c**, At H500, Pt-rich phase recurs at GBs; the composition profile shows Pt enrichment and Ag depletion. **d**, Chemical heterogeneity quantified by the Pearson coefficient μ across all investigated thermal states. Open circles show the average μ across Ag, Au, Pd and Pt for the two starting states, decreasing from 0.49 in the as-deposited state to 0.14 after homogenization.

To determine whether the experimentally reconstructed pathway is consistent with the underlying thermodynamic tendency, we performed hybrid molecular dynamics/Monte Carlo (MD/MC) simulations using a machine-learning interatomic potential (Fig. 4b, Methods). Starting from a chemically random fcc solid solution, the predicted room-temperature equilibrium state separates into a disordered binary Pt-rich phase containing 68 at% Pt and 32 at% Pd and a surrounding matrix, whereas at 700°C a single-phase solid solution is predicted. The latter corresponds to the experimentally homogenized state at 700°C (Fig. 4a), while the low-temperature prediction independently identifies the same Pt-rich phase-selection direction observed experimentally.

Experimentally, the Pt-rich phase at 600°C contains ~40 at% Pt, ~55 at Pd and ~5 at% of Ag and Au combined (Fig. 2g and Extended Data Fig. 6), and continues to evolve compositionally, as evidenced by the progressive Pt enrichment from 500 to 600°C (Fig. 2g). Quantitative compositional agreement with the predicted room-temperature equilibrium phase is therefore neither expected nor required. Instead, experiment and simulation converge on the same direction of phase selection: from a homogeneous solid solution at high temperature towards formation of a Pt-rich phase at lower temperature.

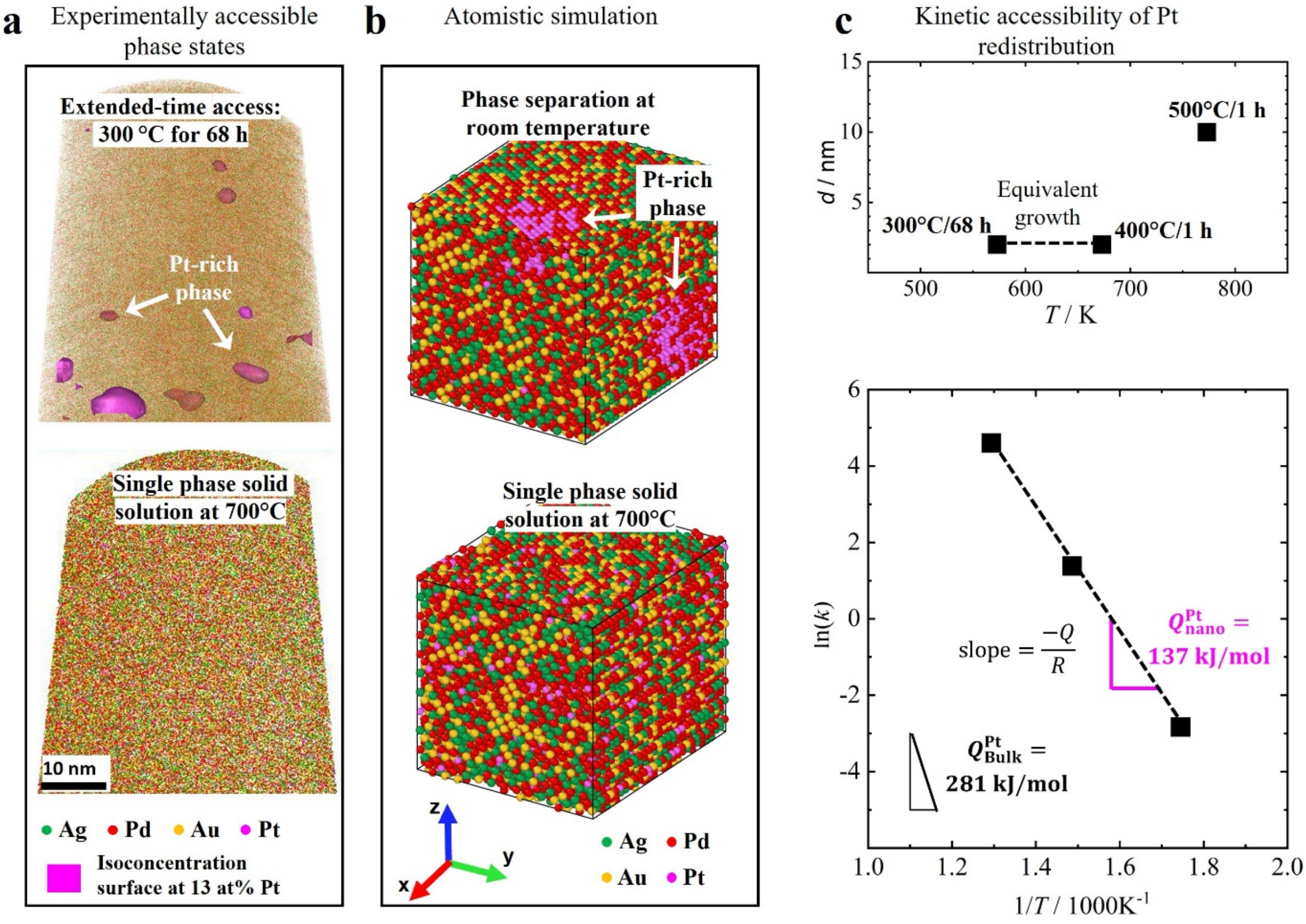

**Fig. 4 Accessing the low-temperature transformation pathway**. **a**, Experimentally accessible phase states. Extended annealing for 68 h reveals Pt-rich phase formation at 300°C, whereas annealing at 700°C produces a homogeneous single-phase solid solution. Pink isoconcentration surfaces correspond to 13 at% Pt. **b**, Representative equilibrium configurations from atomistic simulations at room temperature and 700°C (see text and Methods for details). **c**, Kinetic analysis of Pt redistribution. Comparable Pt-rich particle sizes $d$ after annealing at 300°C for 68 h and 400°C for 1 h show that extended annealing enables comparable redistribution at lower temperature. Arrhenius analysis using $k=d^2/t$ yields an apparent activation energy of 137 kJ/mol, substantially below the reported bulk Pt self-diffusion activation energy of approximately 281 kJ/mol[43].

To quantify the kinetic accessibility of the reconstructed pathway, we analyzed the Pt redistribution across the 300°C, 400°C and 500°C annealing conditions (Fig. 4c). Taking the characteristic Pt-rich particle size $d$ as an effective redistribution length and the annealing time $t$, we define an effective kinetic coefficient $k = d^2/t$. The resulting Arrhenius dependence yields an apparent activation energy of approximately 137 kJ/mol, about half the reported bulk Pt self-diffusion value of ~281 kJ/mol[43]. Based on this bulk Pt self-diffusion kinetics, comparable redistribution would require years to decades (Methods), whereas the nanolabs render it accessible within hours to days. Pure Pt self-diffusion therefore provides an order-of-magnitude benchmark for the observed kinetic acceleration. The lower apparent barrier, the preferential nucleation of the Pt-rich phase at GBs (Fig. 2) and the delayed formation following homogenization (Fig. 3) are consistent with a substantial contribution from defect-assisted redistribution. Because homogenization simultaneously changes both initial chemical heterogeneity and GB density, the value of 137 kJ/mol should therefore be interpreted as an apparent kinetic scale for the overall transformation, not as a measurement of a unique diffusion mechanism.

Across annealing temperature, starting state and time, the Pt-rich phase repeatedly emerges, while the conditions required for its observation change with kinetic accessibility. Its formation shifts from 400°C to 500°C within the 1-h window after homogenization, whereas prolonged annealing reveals the same phase already at 300°C. Thus, the temperature at which the Pt-rich phase first becomes detectable within a fixed annealing time does not necessarily represent a thermodynamic phase boundary. These results, supported by atomistic simulation, demonstrate that our nanolab metallurgy provides experimental access to thermodynamically driven phase evolution that would otherwise remain hidden by kinetic constraints.

Earlier work[26] established that confined nanocrystalline volumes can accelerate otherwise inaccessible phase decomposition. Here, this capability is used to reconstruct a pathway towards lower-free-energy states across temperature, microstructural starting state and time, and to connect recurrent Pt-rich phase formation with independent atomistic predictions. The advance is therefore not accelerated transformation alone, but experimental reconstruction of the pathway connecting a metastable starting state with its low-temperature phase-selection tendency. Neither the as-prepared state nor the predicted equilibrium state alone describes how a real, history-dependent material evolves between them.

More broadly, the long-term stability of metastable CCMs should be considered in terms of the pathways accessible under relevant combinations of temperature, time and microstructural history. A material that appears stable within a conventional experimental window may retain a thermodynamic tendency towards transformation that becomes accessible over longer times or as its microstructure evolves. In the investigated alloy, phase evolution is accompanied by an approximately 3.7-fold loss in HER activity. Although HER activity is governed primarily by the surface, formation of Pt-rich phase strongly depletes Pt from the surrounding matrix and is therefore expected to reduce the Pt available at surface regions outside the Pt-rich phase. This interpretation is consistent with the intrinsically high HER activity of Pt relative to the other constituent elements[44] and with the SECCM library map, where Pt-richer compositions exhibit higher activity (Extended Data Fig. 3). This illustrates how kinetically delayed phase evolution can substantially alter the functional properties of a material that appears stable within conventional experimental windows.

**Summary**

We reveal a hidden low-temperature transformation pathway in a metastable CCM by systematically varying annealing temperature, time, and microstructural starting state. Recurrent formation of a Pt-rich phase across distinct kinetic conditions and microstructural histories, together with independent atomistic predictions, identifies a common thermodynamic tendency whose experimental accessibility is governed by kinetics. Nanolab metallurgy makes this otherwise inaccessible evolution experimentally resolvable and shows that the apparent onset of transformation can be a kinetic observation boundary rather than a thermodynamic phase boundary. The associated loss of electrocatalytic activity illustrates how delayed phase evolution can alter functional performance. Revealing such pathways bridges the gap between metastable as-synthesized states and predicted equilibrium states, providing an experimental basis for designing metastable materials for long-term phase stability and functional persistence.

## Methods

### Thermodynamic descriptors for composition selection

The Ω–δ criteria were used as empirical thermodynamic descriptors to assess the phase-forming tendency of the Ag-Au-Pd-Pt model alloy, following refs..[45,46] The descriptors were calculated as $\Omega = \frac{T_m \Delta S_{mix}}{|\Delta H_{mix}|}$ and $\delta = 100\sqrt{[\Sigma c_i (1 - \frac{r_i}{\bar{r}})^2]}$, where $T_m$ is the composition-weighted melting temperature, $\Delta S_{mix}$ is the ideal configurational entropy of mixing, $\Delta H_{mix}$ is estimated from the binary mixing enthalpies of the constituent pairs, $c_i$ and $r_i$ are the atomic fraction and atomic radius of element $i$, respectively, and $\bar{r}$ is the composition-weighted average atomic radius. The atomic radii, melting temperatures, and the binary mixing enthalpies used in the calculations were taken from refs.[47–49], respectively. The resulting Ω and δ values locate $Ag_{24}Au_{20}Pd_{50}Pt_6$ near the empirical transition between solid-solution and phase-separation regimes (Extended Data Fig. 4), supporting its selection for investigating metastable phase evolution.

### Materials library fabrication and high-throughput characterisation by EDX and SECCM

The Ag-Au-Pd-Pt combinatorial thin-film materials library (ML) was prepared by magnetron co-sputtering onto a 100-mm-diameter c-plane sapphire wafer using an eight-cathode ATC-2200-UHV system (AJA International). Four elemental sources were operated simultaneously to generate continuous lateral composition gradients across the substrate. Elemental targets (50.8 mm diameter) of Ag (99.99%, Testbourne), Au (99.99%, Testbourne), Pd (99.95%, Testbourne) and Pt (99.99%, Evochem) were used. Before deposition, the substrate was cleaned in situ for 10 min using a 25 W RF bias in Ar (4Pa, 40 sccm), and the targets were pre-sputtered for 60 s at 50 W behind closed shutters under 4 Pa Ar. Deposition was performed at room temperature in high-purity Ar (99.9999%) at 0.67 Pa and 30 sccm, with a base pressure of approximately $3.5 \times 10^{-7}$ Pa and a target–substrate distance of 178 mm. Deposition rates were calibrated as a function of sputtering power using an in situ quartz-crystal monitor. The library was deposited for 597 s using 23 W DC (Ag), 52 W RF (Au), 70 W DC (Pd) and 36 W RF (Pt), yielding a nominal film thickness of approximately 90 nm at the wafer centre. The ML was divided into 342 uniformly distributed measurement areas (MAs; $4.5 \times 4.5$ mm$^2$) using a predefined wafer grid for subsequent compositional and electrochemical characterization.

The as-deposited composition was determined at the predefined MAs by automated energy-dispersive X-ray spectroscopy (EDX) using a scanning electron microscope (TESCAN VEGA 3) equipped with a silicon-drift EDX detector (Bruker QUANTAX XFlash 7). Measurements were performed at 20 kV, a working distance of 15 mm and a field of view of 300 μm. The beam current was adjusted to approximately 100 kcps at ~20% detector dead time, and spectra were acquired to $1 \times 10^6$ counts. The resulting compositions were used to construct the elemental composition maps of the materials library. The composition $Ag_{24}Au_{20}Pd_{50}Pt_6$ was selected for the nanolab experiments based on SECCM screening (Extended Data Fig. 3). and thermodynamic assessment using the Ω–δ descriptors (Extended Data Fig. 4), described below.

**Scanning electrochemical cell microscopy**
Hydrogen evolution reaction (HER) activity of the Ag-Au-Pd-Pt materials library was measured in the as-deposited and annealed (600°C for 1 h) states using a custom-built long-range scanning electrochemical cell microscopy (SECCM) system operated in hopping mode, as described previously[50]. Single-barrel quartz capillaries (QF-120-90-10, Science Products) were laser-pulled to an opening diameter of approximately 500 nm and filled with 0.1 M $HClO_4$ containing 0.1 M $LiClO_4$ (pH ≈ 1.2), with an Ag/AgCl (3 M KCl) quasi-reference counter electrode. At each measurement position, meniscus contact defined a confined electrochemical cell. Three conditioning cyclic voltammograms were recorded at 5 V $s^{-1}$, followed by a HER linear sweep voltammogram at 1 V $s^{-1}$. Repeated measurements within each composition region were averaged. Potentials were converted to the reversible hydrogen electrode (RHE) scale according to $E_{RHE} = E_{applied} + E_{Ag/AgCl} + 0.059$ pH, with $E_{Ag/AgCl} = 0.210$ V for Ag/AgCl (3 M KCl). HER activity maps were constructed from the current density at −300 mV versus RHE, where a pronounced HER response enabled robust comparison across the as-deposited and annealed materials libraries (Extended Data Fig. 3).

**Nanolab fabrication and thermal treatment**
The $Ag_{24}Au_{20}Pd_{50}Pt_6$ composition was selected from the Ag-Au-Pd-Pt materials library by combining thermodynamic and functional considerations. The composition lies within the transitional solid-solution plus intermetallic (S+I) region of the empirical Ω-δ phase-selection map (Extended Data Fig. 4), making it suitable for investigating the evolution of a metastable solid solution towards lower-free-energy states. SECCM screening of the as-deposited and annealed library provided an additional functional basis for selecting this composition (Extended Data Fig. 3).

The selected $Ag_{24}Au_{20}Pd_{50}Pt_6$ composition was deposited onto $SiO_2$/Si microtip arrays prepared from commercially available presharpened Si microtip arrays (MicroTip™, CAMECA Instruments), each containing 36 tips. The Si arrays were thermally oxidized by Team Nanotec in a horizontal furnace (CPS382-II, Centrotherm) under an oxygen atmosphere at 1080°C to form an approximately 22-nm-thick thermal $SiO_2$ barrier[51], which suppresses intermixing between the alloy film and the Si substrate during subsequent thermal treatments[30]. Ag, Au, Pd and Pt were magnetron co-sputtered in a physical vapor deposition system (DCA, Finland) equipped with five magnetron cathodes in a sputter-down configuration. The microtip arrays were positioned near the confocal point of the cathodes at a target-to-substrate distance of approximately 185 mm. Before deposition, the targets were pre-sputtered for 300 s behind closed shutters. Deposition was performed at a substrate-holder temperature of 25°C and an Ar pressure of 0.67 Pa, with a chamber base pressure of $3.4 \times 10^{-5}$ Pa. The substrate table was rotated at 10 revolutions per minute (rpm) during co-deposition to obtain a uniform alloy composition across the microtip array. The sputter powers were 35 W (DC) for Ag, 112 W (RF) for Au, 300 W (RF) for Pd and 34 W (DC) for Pt with a deposition time of 132 s. The resulting film thickness was approximately 70-80 nm, as calibrated on a planar Si reference substrate. Each coated 36-tip microtip array constitutes a *nanolab* comprising 36 nominally equivalent nanoscale alloy volumes that serve as individual APT specimens.

To investigate the phase-evolution pathway, three complementary probes varying annealing temperature, microstructural starting state and time were employed (Extended Data Fig. 5). For the temperature probe, a nanolab comprising 36 APT specimens was subjected to sequential 1 h annealing treatments under vacuum ($8.6 \times 10^{-5}$ Pa) from 100°C to 700°C in 100°C increments, with selected specimens analysed by APT after each treatment. The state homogenized at 700°C in the same nanolab was then used for the microstructural starting-state probe and subsequently annealed from 100 to 500°C in 100°C increments. For the time probe, a separate nanolab was annealed at 300°C for progressively extended durations, with intermediate APT analyses of selected specimens, reaching a cumulative annealing time of 68 h.

Morphological stability of the nanolabs during high-temperature annealing was assessed by comparing the alloy film at the tip apex with the film on the surrounding shank and planar substrate. After annealing at 700°C for 1 h, the thin film on the flatter surrounding regions dewets and breaks into isolated features, whereas the alloy volume at the cone-shaped tip apex retains its morphology (Extended Data Fig. 10). This observation establishes that the nanolab geometry suppresses dewetting under the highest-temperature treatment used in this study and preserves the confined alloy volume required for subsequent APT analysis.

**Assessment of APT-specimen curvature effects on thermodynamic phase selection**

The possible influence of tip curvature on the chemical potential was estimated using the Gibbs-Thomson relation. For a tip of radius $r$, the curvature-induced chemical-potential shift was estimated as $\Delta\mu = 2\gamma V_m / r$, where $\gamma$ is the surface energy and $V_m$ is the molar volume. The elemental compositions, molar volumes and literature surface energies[52–54] used for the estimate are listed in Extended Data Table 1.

To obtain a conservative upper-bound estimate for Pt redistribution, the surface energy and molar volume of pure Pt were used, because Pt has the highest surface energy among the constituent elements considered. For a representative tip radius of approximately 30 nm, the resulting curvature-induced chemical-potential shift is approximately 1.5 kJ/ mol (0.016 eV). Although finite relative to the thermal energy scale, this contribution remains small at the tip dimensions investigated and decreases further with increasing radius; substantial curvature-induced shifts emerge only for radii of several nanometers (Extended Data Fig. 11). Thus, capillary effects are unlikely to substantially alter the intrinsic phase-selection tendency under the present experimental conditions.

**Near-atomic-scale structural and compositional characterization**

Following the respective annealing treatments, individual nanolabs were analyzed by atom probe tomography (APT) to characterize phase evolution as a function of starting state as well as annealing temperature and time. APT measurements were performed using a LEAP5000XR (CAMECA Instruments) in laser-pulsing mode at 60 K, with a laser pulse energy of 60 pJ, a pulse repetition rate of 125 kHz and a target detection rate of 0.005 atoms per pulse. Data reconstruction and analysis were performed using IVAS 3.8.16 (CAMECA Instruments), following established reconstruction and analysis procedures[55,56].

Chemical heterogeneity was quantified by frequency-distribution analysis[42] using the Pearson coefficient , which measures the deviation of the experimental elemental distribution from that expected for a random binomial distribution. Lower values therefore indicate distributions closer to random, whereas increasing indicates increasing chemical heterogeneity.

For semi-correlative TEM characterization, selected nanolabs were transferred from the microtip arrays to TEM grids using a focused ion beam (FIB) system (Helios G4 CX, Thermo Fisher Scientific), following the procedure described previously[41]. Specimens were transferred and thinned at 30 kV, followed by final milling at 8 kV to minimize ion-beam-induced damage. TEM characterization was performed using an aberration-corrected JEM-ARM200F (JEOL) operated at 200 kV. Elemental distributions were characterized by energy-dispersive X-ray spectroscopy (EDX), with compositions quantified from normalized detector counts. Nano-beam diffraction for orientation mapping was performed using DigiStar P2000 ASTAR system (NanoMEGAS). Diffraction patterns were acquired at 200 kV, using 1 nm probe at camera length of 50 cm, with a 2.5 nm step size. The resulting diffraction index map, and inverse pole figure (IPF) were generated and analyzed using NanoMEGAS MapViewer 2.0.8.389 software.

**Binomial frequency distribution analysis**

Chemical heterogeneity in the APT datasets was quantified using binomial frequency distribution analysis following Moody *et al.*.[42] Each three-dimensional APT dataset was partitioned into $N$ discrete blocks containing an equal number of atoms, $n_b$=100. For each element $A$, the number of atoms $n$ occurring in each block was counted and the experimental frequency distribution was compared with the corresponding binomial distribution expected for a random distribution of atoms. The theoretical frequency is $f(n) = N\binom{n_b}{n}c_A^n(1 - c_\mathrm{A})^{n_b-n}$, where $c_\mathrm{A}$ is the overall atomic fraction of element $A$. The departure of the experimentally measured distribution $e(n)$ from the corresponding binomial distribution was quantified using the chi-square statistic $\chi^2 = \sum_{n=0}^{n_b} \frac{[e(n)-f(n)]^2}{f(n)}$. Because $\chi^2$ depends on the number of sampled blocks, the Pearson coefficient of contingency, $\mu = \sqrt{\frac{\chi^2}{N+\chi^2}}$, was used to facilitate comparison among datasets of different sizes, following the method of Moody *et al.*.[42]

The Pearson coefficient $\mu$ ranges from 0 to 1, where $\mu$ approaching 0 indicates close agreement with the random binomial distribution and increasing $\mu$ indicates an increasing departure from randomness. Because $\mu$ quantifies the magnitude of this departure but does not by itself identify its spatial origin, the values were interpreted together with the corresponding APT reconstructions and composition profiles. For each thermal state in Fig. 3d, $\mu$ was calculated separately for Ag, Au, Pd and Pt; the open circles represent the arithmetic mean of the four element-specific $\mu$ values. The analysis was applied to the as-deposited and sequentially annealed states, the 700°C-homogenized state, and the subsequently annealed H400 and H500 states to quantify the evolution of chemical heterogeneity associated with phase decomposition, homogenization and re-formation of the Pt-rich phase.

**Hybrid molecular dynamics and Monte Carlo simulations**

All atomistic simulations were performed using LAMMPS software[57] and the Graph Atomic Cluster Expansion (GRACE/FS) potential[58] to describe interatomic interactions in the Ag-Au-Pd-Pt system. The simulation cell contained 23,520 atoms with an initial random distribution. Periodic boundary conditions were applied in all directions. The system was first equilibrated at the target temperature (300K or 973K, corresponding to room temperature and 700°C, respectively) and zero pressure in the isothermal-isobaric (NPT) ensemble for 50 ps to correctly adjust the volume.

Subsequently, a hybrid molecular dynamics/Monte Carlo (MD/MC) procedure was employed to reach thermodynamic equilibrium. Molecular dynamics simulations were performed in the canonical ensemble (NVT) at the target temperature, the equations of motions were integrated using the velocity Verlet algorithm with a timestep of 1 fs for a total of 1 ns. Every 1 ps a 1000 Monte Carlo exchanges were attempted between the four different chemical elements, using the "fix atom/swap" command from the MC package[59] as implemented in LAMMPS. The number of atoms of each species was constant throughout the MD/MC procedure. Temperature and pressure were controlled via the Nosé-Hoover thermostat and barostat[60,61], with damping parameters of 0.1 ps and 2 ps, respectively. The simulation trajectories were visualized using OVITO[62].

**Estimation of the apparent activation energy for Pt redistribution**

The apparent activation energy for Pt redistribution was estimated from the temperature- and time-dependent evolution of Pt-rich phase observed by APT. Pt was used as the kinetic tracer because it is the slowest-diffusing constituent among Ag, Au, Pd and Pt. Assuming diffusion-controlled redistribution, the characteristic redistribution distance d follows $d^2 \propto Dt$, where d is the characteristic particle size used as a measure of redistribution distance, $D$ is the effective diffusivity and t is the annealing time. The temperature dependence of $D$ was described by the Arrhenius relation $D = D_0\, exp(-Q/\mathrm{R}T)$, where $D_0$ is the pre-exponential factor, $Q$ is the apparent activation energy, R is the gas constant and T is the absolute temperature.

Particle-growth analysis: APT reconstructions after 1 h at 400°C and 500°C show Pt-rich phase formation associated with GBs. Their number density remains approximately constant while their characteristic size increases substantially, consistent with growth of pre-existing Pt-rich phase rather than classical Ostwald ripening involving extensive particle dissolution and reprecipitation. Because the annealing time was identical at both temperatures ($t_{400} = t_{500} = 1$ h), the effective diffusivity ratio can be expressed as $D_{500}/D_{400} = {d_{500}}^2/{d_{400}}^2$, where $D_{400}$ and $D_{500}$ are the effective diffusivities and $d_{400}$ and $d_{500}$ are the characteristic particle sizes after annealing at $T_{400} = 673$ K (400°C) and $T_{500} = 773$ K (500°C), respectively. Combining this relation with the Arrhenius expression gives $Q = R\, ln({d_{500}}^2/{d_{400}}^2)/[(1/T_{400}) - (1/T_{500})]$. Using the experimentally measured particle-size ratio $d_{500}/d_{400} \approx 5$ yields $Q \approx 139$ kJ/ mol.

Time-temperature equivalence: An independent estimate was obtained by comparing Pt-rich phase formed after 68 h at 300°C and 1 h at 400°C. Their comparable characteristic dimensions suggest comparable Pt redistribution distances. Here, $T_{300} = 573$ K and $T_{400} = 673$ K are the absolute annealing temperatures, and $t_{300} = 68$ h and $t_{400} = 1\,h$ are the corresponding annealing times. Assuming equivalent redistribution distances, $D_{300}t_{300} \approx D_{400}t_{400}$, where $D_{300}$ and $D_{400}$ are the effective diffusivities at 300 and 400°C, respectively. Combining this relation with the Arrhenius expression gives $Q = \mathrm{R}\, ln(t_{300}/t_{400})/[(1/T_{300}) - (1/T_{400})]$. Using the experimental annealing times gives $Q \approx 135$ kJ/mol.

Arrhenius representation: For the Arrhenius representation in Fig. 4c, the characteristic particle size was converted to the time-normalized kinetic coefficient $k = d^2/t$, where k serves as a relative measure of the effective redistribution kinetics and is proportional to $D$ under the diffusion-controlled-growth approximation. Linear regression of $ln(k)$ against $1/T$ yields an apparent activation energy of approximately 137 kJ/mol, consistent with the independent particle growth and time-temperature estimates. The reported value is therefore treated as an apparent kinetic parameter for the overall Pt-redistribution process rather than a direct measurement of a single tracer-diffusion mechanism.

**Order-of-magnitude comparison with bulk Pt lattice diffusion**

Characteristic redistribution times for conventional lattice diffusion were estimated using literature parameters for Pt self-diffusion in coarse-grained bulk Pt. The diffusion coefficient was calculated from $D = D_0 \exp(-Q/\mathrm{R}T)$. Literature values of $D_0 = 5 \times 10^{-5}$ m$^2$/s and $Q$ = 281 kJ/mol were used[43]. The characteristic redistribution time was estimated as $t \approx x^2/D$, where $x$ is the characteristic redistribution distance. Using $x = 2$ nm for the Pt-rich phase observed at 400°C gives a characteristic redistribution time of approximately 47 years, whereas decomposition is experimentally observed after 1 h. Using $x$=10 nm for the larger Pt-rich particles observed at 500°C gives approximately 1.5 years, compared with an experimental annealing time of 1 h. These estimates are intended as order-of-magnitude references based on pure coarse-grained Pt self-diffusion rather than quantitative predictions for the multicomponent alloy. They therefore provide a baseline for comparing conventional lattice diffusion timescales with the experimentally observed redistribution kinetics in the nanoscale volume.

**Acknowledgements** This research was funded by the Deutsche Forschungsgemeinschaft (DFG, German Research Foundation) SFB 1625, project number 506711657. Y. L. and E. A. acknowledge support through subproject B01. Q. B., M. M. and R. D. acknowledge support through subproject A03. A. K. acknowledges support through subproject S. A.L. and N.P. acknowledge support through subprojects A01 abd A02. S. R.Z. acknowledges support by the early career research group of the SFB 1625, as well as subprojects A01 and C01 (Prof. Dr. Wolfgang Schuhmann). B. G. acknowledges support through subproject B04. D. R. acknowledges support through subproject A02. The authors acknowledge ZGH, Ruhr University Bochum for access to APT, TEM, FIB, SEM, and SECCM; and Dr. Janine Pfetzing-Micklich through Subproject S for support and coordination of the use of the ZGH facilities.

**Author contributions**

Y.L. and A.L. conceived the project and designed the research. Y.L. was the leading research scientist of this work who carried out all APT analyses and analyzed all data. E.A. fabricated the nanolab specimens and performed the heat treatments. Q.B. performed the simulations. A.K. performed the TEM analysis. N.P. synthesized and characterized the materials library, including EDX mapping. R.Z. performed the SECCM measurements of the materials library. A.S. supported the materials synthesis. M.M. and R.D. supervised the simulations. Y.L., B.G, D.R. and A.L. discussed and rationalized the observations. Y.L. wrote the initial manuscript. A.L. supported writing of the paper. All authors reviewed and approved the final version of the paper.

# Extended data

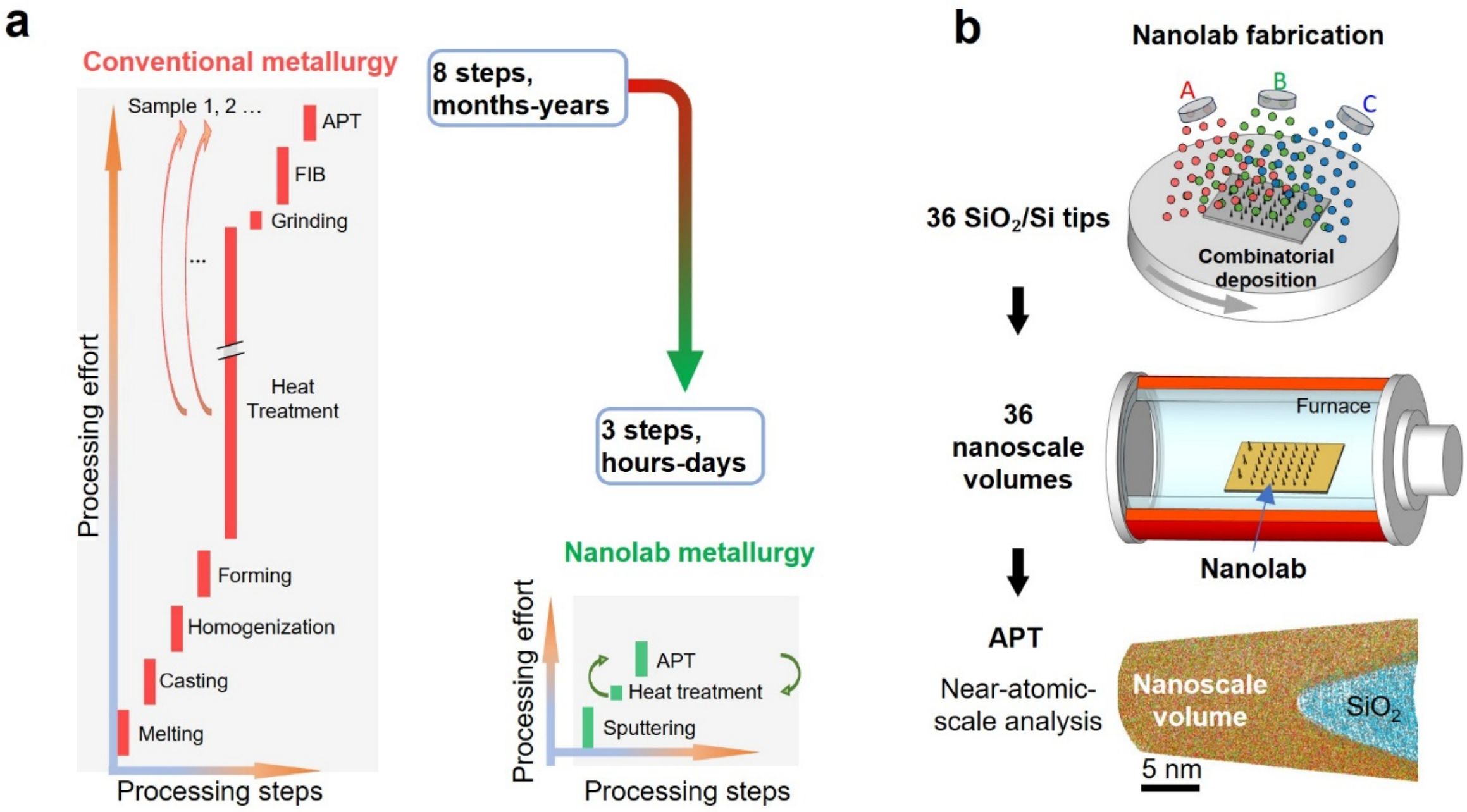

**Extended Data Fig. 1 Nanolab metallurgy and nanolab fabrication**. **a**, Schematic comparison of conventional and nanolab metallurgy. Conventional workflows involve multiple bulk-processing and specimen-preparation steps, including FIB-based specimen preparation for APT, whereas nanolab metallurgy integrates thin-film deposition, heat treatment and direct APT analysis. Short redistribution distances and defect-assisted transport accelerate atomic redistribution[26], enabling metallurgical evolution to be investigated on hours-to-days rather than months-to-years timescales. **b**, Nanolab fabrication and analysis. Commercially available presharpened Si microtip arrays (MicroTip$^{TM}$, CAMECA Instruments), each containing 36 tips, were thermally oxidized by Team Nanotec to form $SiO_2$/Si microtip arrays for combinatorial deposition. The resulting 36 nanoscale alloy volumes serve as individual APT specimens, with the coated 36-tip array constituting a *nanolab*. Repeated heat-treatment and APT-analysis cycles on selected specimens enable the phase-evolution pathway to be probed through complementary variations in annealing temperature, microstructural starting state and time (Fig. 1b, c; Extended Data Fig. 5).

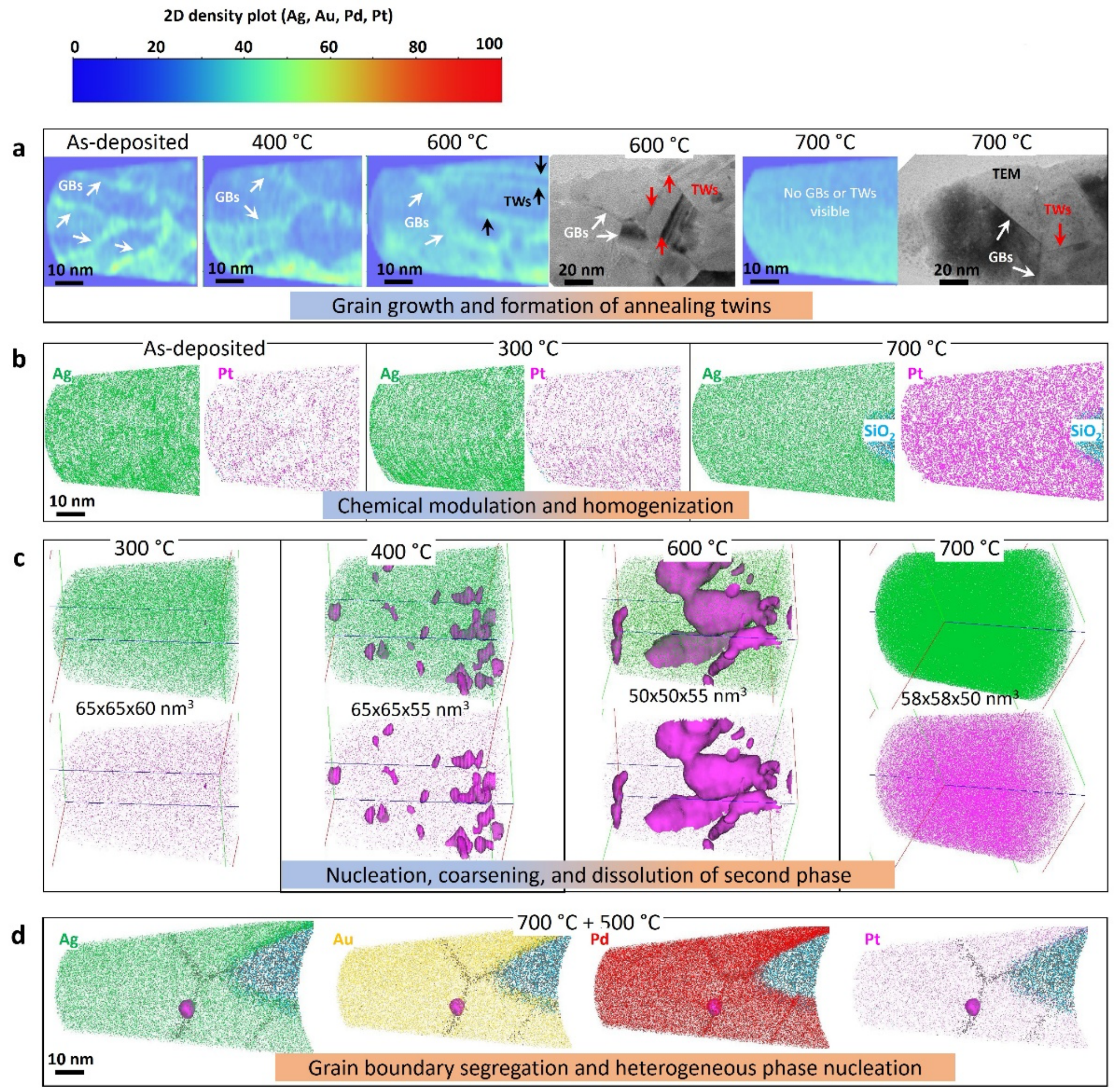

**Extended Data Fig. 2 Representative microstructural evolution demonstrating characteristic metallurgical phenomena within the nanolabs using nanolab metallurgy. a**, Grain growth and formation of annealing twins during heat treatment, revealed by 2D density maps and TEM. **b**, Evolution from the chemically modulated as-deposited state towards homogenization at 700°C, shown by Ag and Pt atom maps. **c**, Formation, coarsening and dissolution of the Pt-rich phase with increasing temperature. Pink isoconcentration surfaces correspond to 13 at% Pt. **d**, Grain-boundary segregation and heterogeneous nucleation of the Pt-rich phase after homogenization at 700°C followed by annealing at 500°C. The 2D density color scale is common to all density maps.

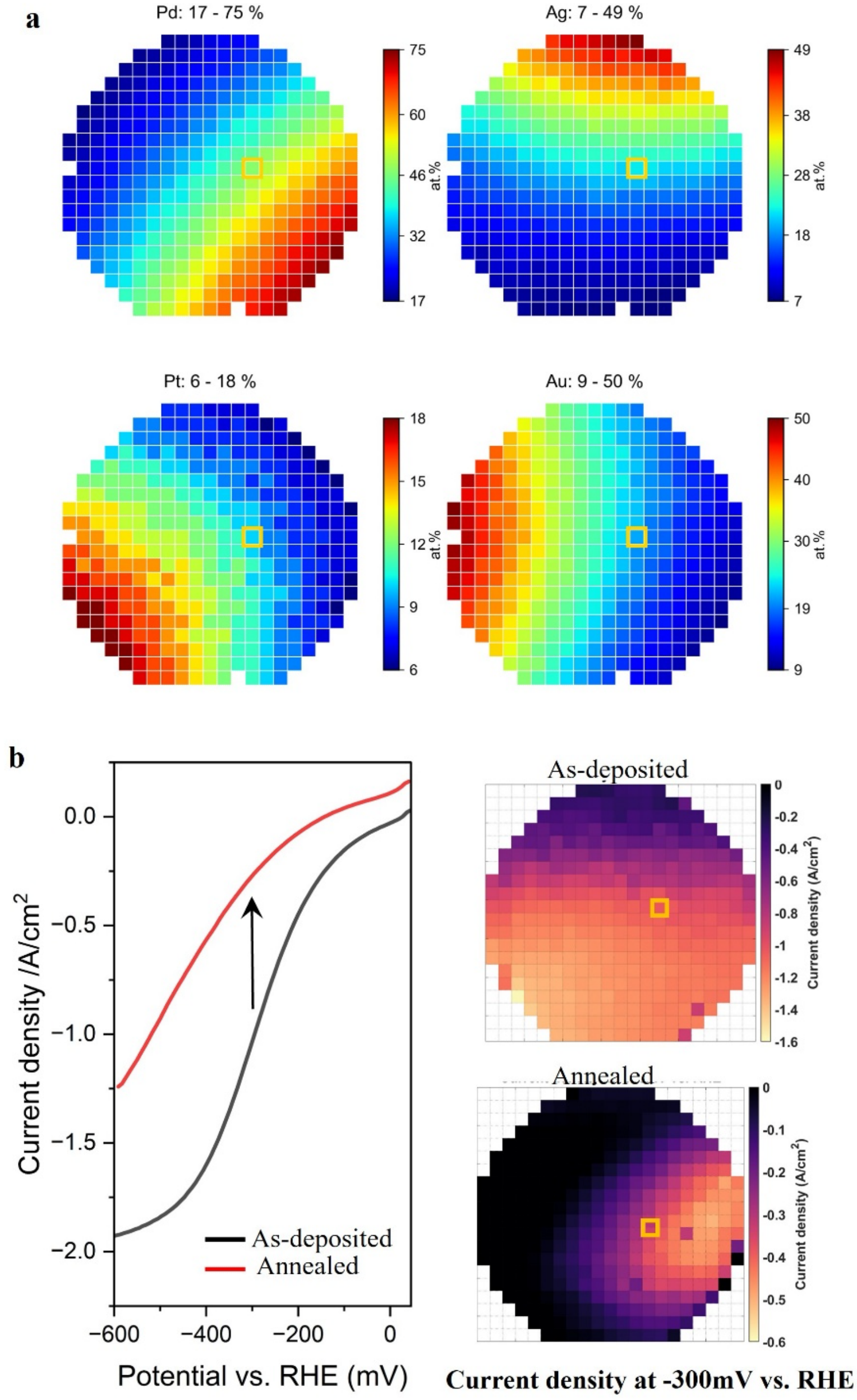

**Extended Data Fig. 3 Selection and HER activity of the Ag-Au-Pd-Pt composition. a,** EDX composition maps of the Ag-Au-Pd-Pt combinatorial materials library showing the spatial distributions of Pd, Ag, Pt and Au. The composition selected for the nanolab experiments, $Ag_{24}Au_{20}Pd_{50}Pt_{6}$, is indicated by the yellow square. b, HER linear sweep voltammograms (LSVs) of the selected composition in the as-deposited state and after annealing to 600°C for 1 h, together with the corresponding SECCM current-density maps at -300 mV versus the reversible hydrogen electrode (RHE). The selected composition is indicated by yellow squares, shows a pronounced decrease in HER current density after annealing.

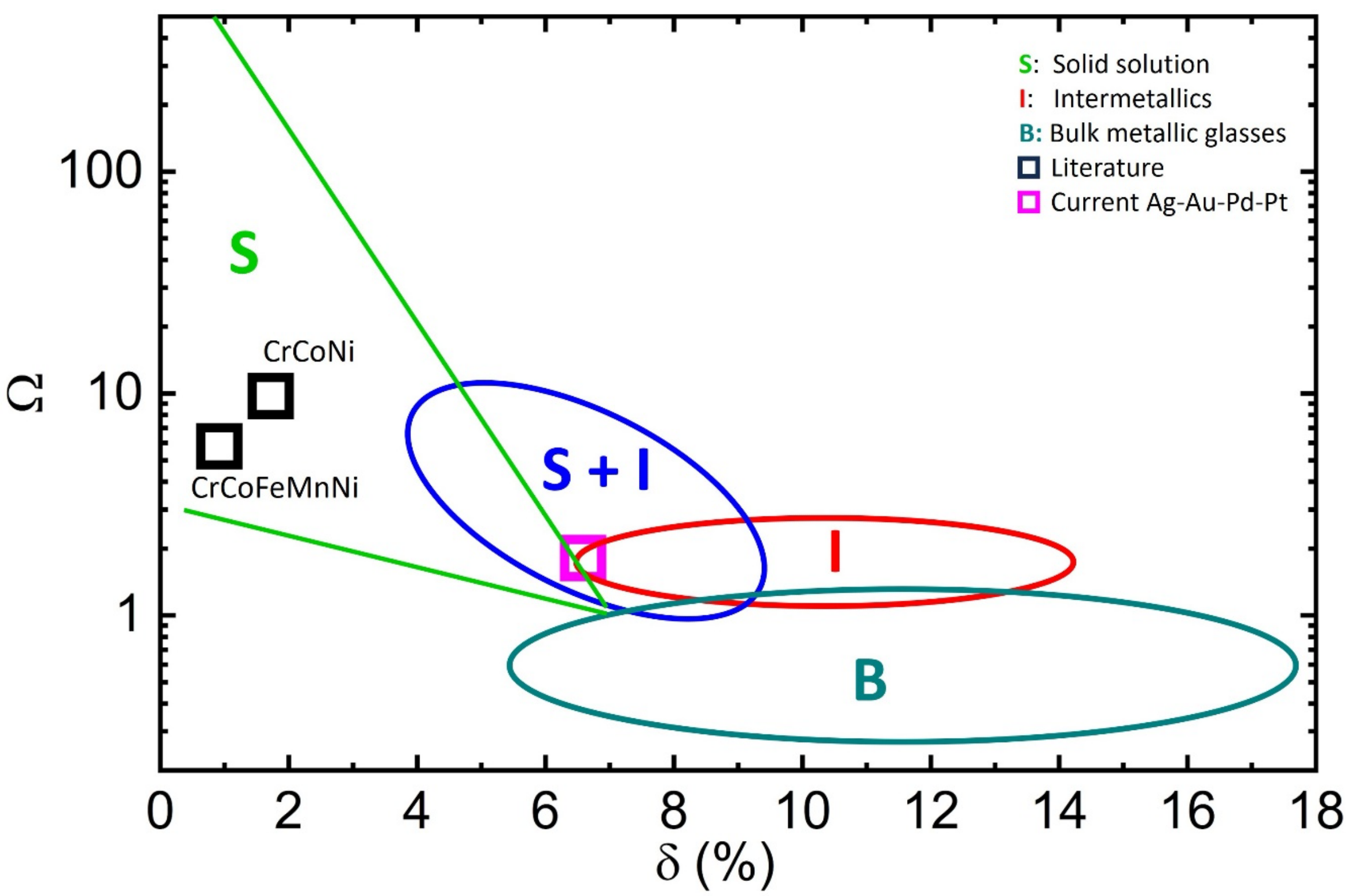

**Extended Data Fig. 4 Ω-δ phase-selection map for the investigated alloy.** Empirical phase-selection regions for solid solutions (S, green), solid solutions plus intermetallics (S+I, blue), intermetallics (I, red) and bulk metallic glasses (B, teal), adapted from Yang and Zhang[45]. The $Ag_{24}Au_{20}Pd_{50}Pt_6$ composition investigated here is indicated by the magenta square; black squares denote CrCoNi and CrCoFeMnNi (Cantor alloy) alloys investigated in our previous work[26,27].

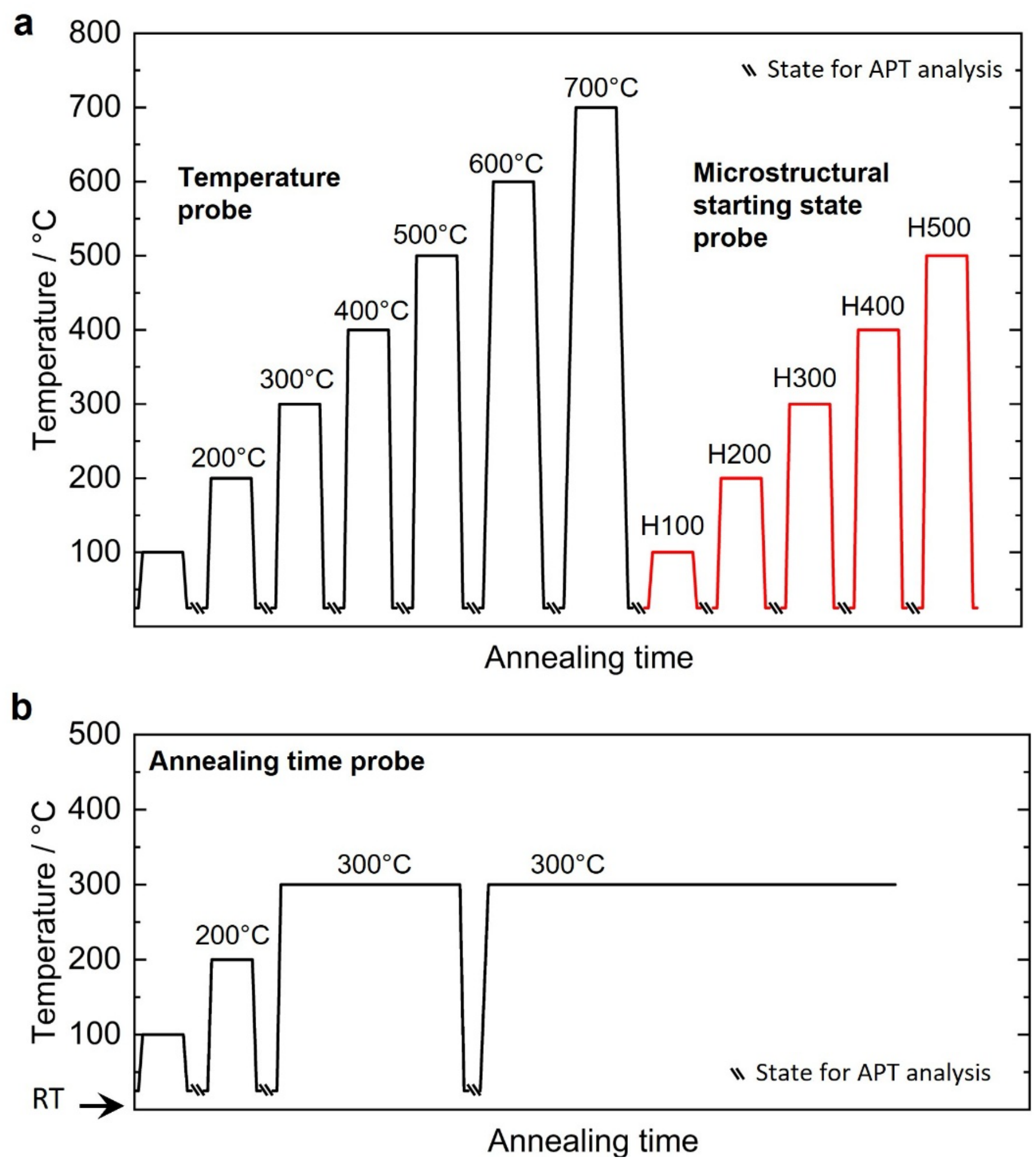

**Extended Data Fig. 5 Schematic annealing procedures for probing hidden phase evolution. a,** Temperature and microstructural starting-state probes. The temperature probe comprises sequential annealing from 100 to 700°C in 100°C increments. The microstructural starting-state probe uses the homogenized state obtained at 700°C as the starting state; H100-H500 represent samples annealed at 100-500°C, respectively, after homogenization at 700°C. **b,** Time probe, comprising initial stepwise annealing to 300°C followed by extended annealing at 300°C.

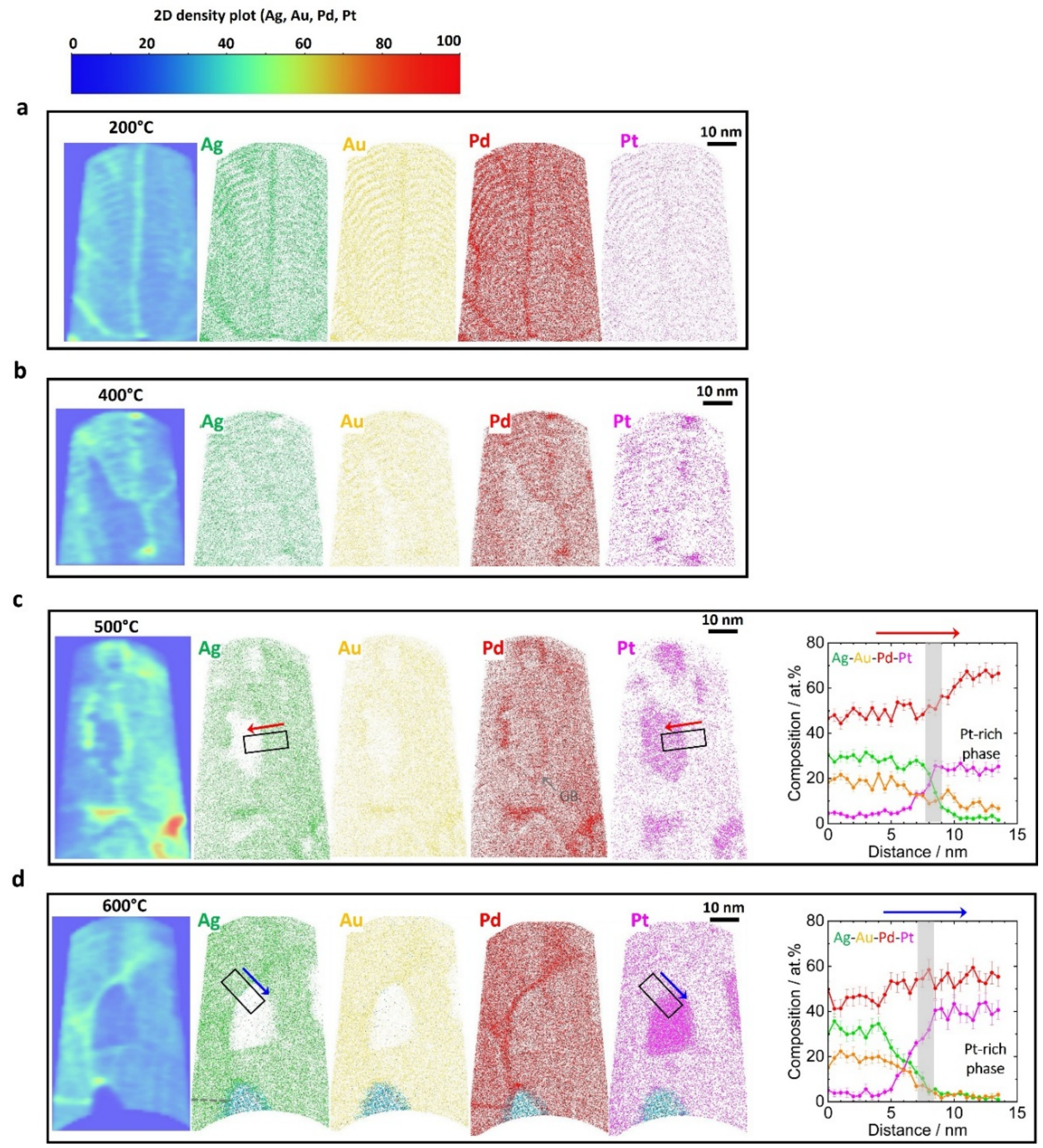

**Extended Data Fig. 6 Full elemental distributions during temperature-dependent phase evolution of metastable $Ag_{24}Au_{20}Pd_{50}Pt_{6}$.** APT reconstructions showing Ag, Au, Pd and Pt distributions after 1 h annealing at **a,** 200°C, **b,** 400°C, **c,** 500°C and **d,** 600°C, complementing the Ag- and Pt-focused maps presented in Fig. 2. Two-dimensional (2D) atomic-density maps are shown alongside the elemental reconstructions to visualize the underlying grain structure. At 200°C, no distinct Pt-rich phase is detected. Pt-rich, Ag-depleted regions emerge at 400°C and subsequently grow and coarsen at 500°C and 600°C. **c,d,** Composition profiles across the matrix-Pt-rich phase interfaces showing the redistribution of all four alloying elements. Rectangles and arrows indicate the regions of interest and compostion profile directions, respectively. Shaded regions denote the interfaces. Further Pt enrichment and Ag depletion are observed from 500°C to 600°C, while Au is also depleted from the Pt-rich phase. The colour scale applies to the two-dimensional atomic-density maps.

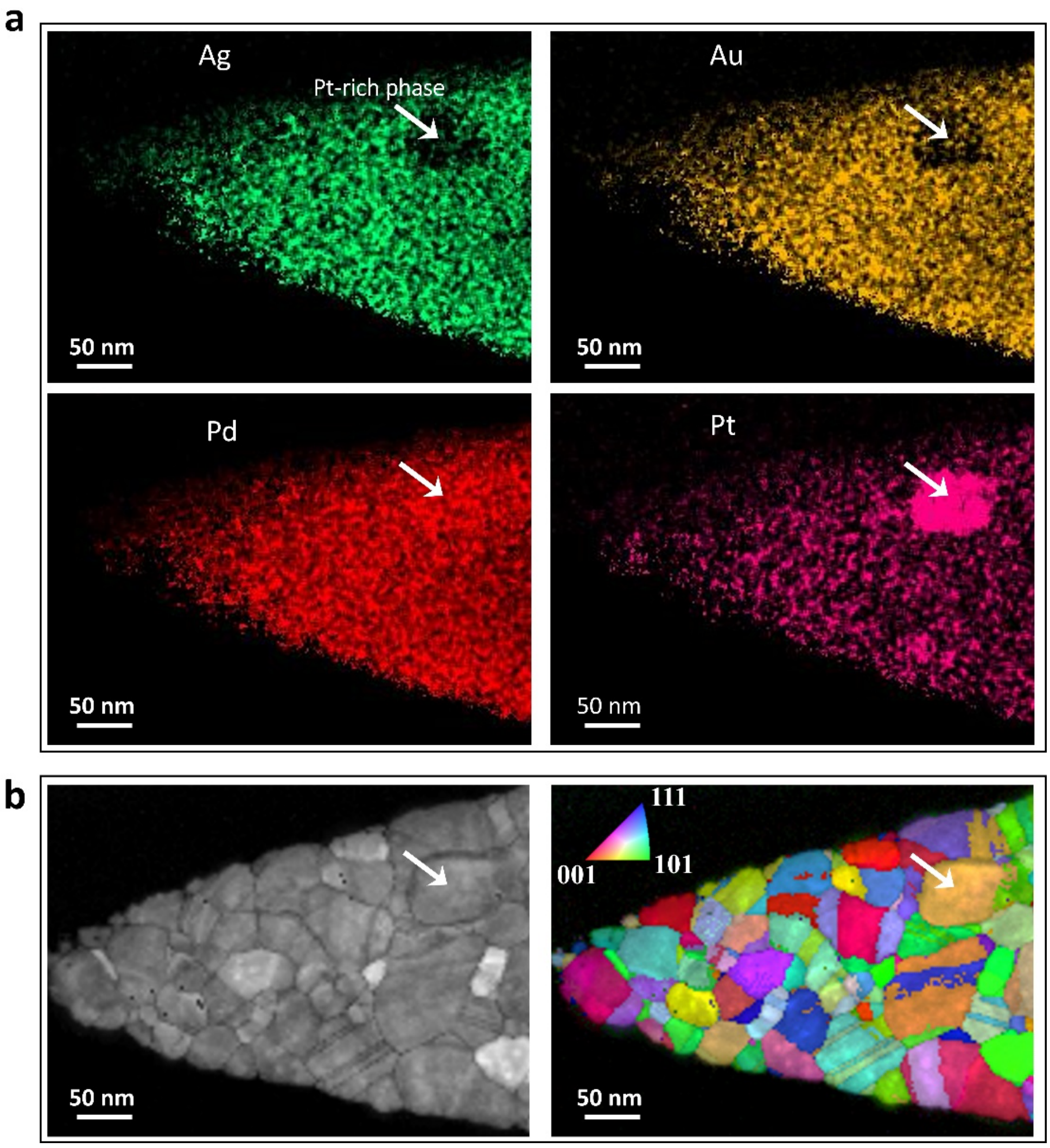

**Extended Data Fig. 7 TEM characterization of the phase constitution comprising a fcc solid solution matrix phase and a Pt-rich phase after annealing at 600°C for 1 h. a,** TEM-EDX elemental maps of Ag, Au, Pd and Pt. Arrows indicate the Pt-rich phase. **b,** ASTAR nanobeam diffraction index map (template matching index) of the same region. Arrows indicate the locations of the Pt-rich phase identified by TEM-EDX. The nanobeam diffraction analysis confirms that the Pt-rich phase is crystalline and has the same fcc crystal structure as the surrounding matrix.

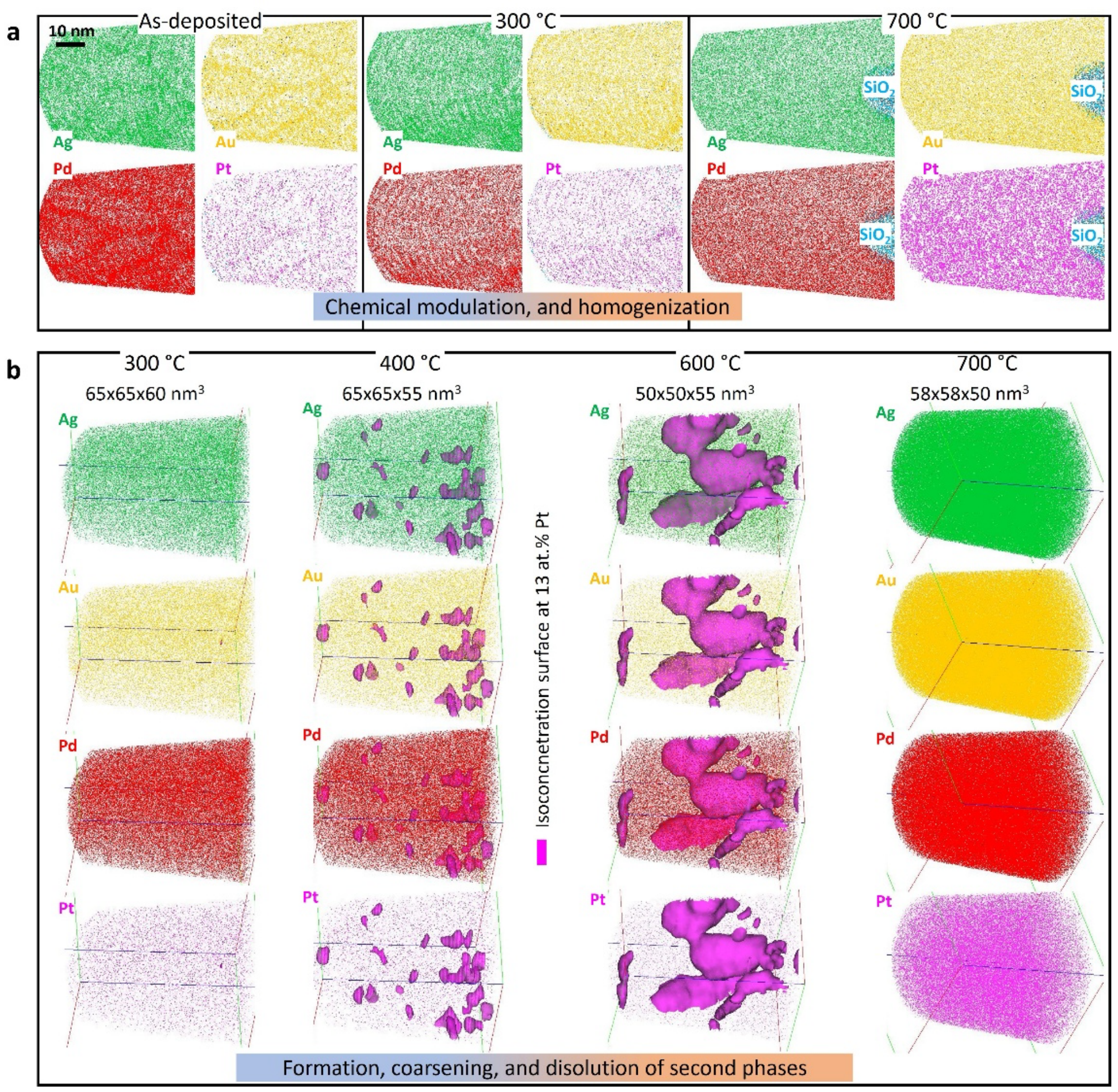

**Extended Data Fig. 8 Temperature-dependent phase evolution using a nanolab. a,** Two-dimensional APT atom maps from 5-nm-thick slices of nanolabs in the as-deposited state and after annealing at 300°C and 700°C for 1 h, showing chemical modulation and subsequent homogenization. **b,** Three-dimensional APT reconstructions after annealing at 300, 400, 600 and 700°C for 1 h. Pt-rich regions are visualized using 13 at% Pt isoconcentration surfaces, revealing the absence of detectable Pt-rich phase at 300°C, its formation and coarsening at intermediate temperatures, and dissolution at 700°C.

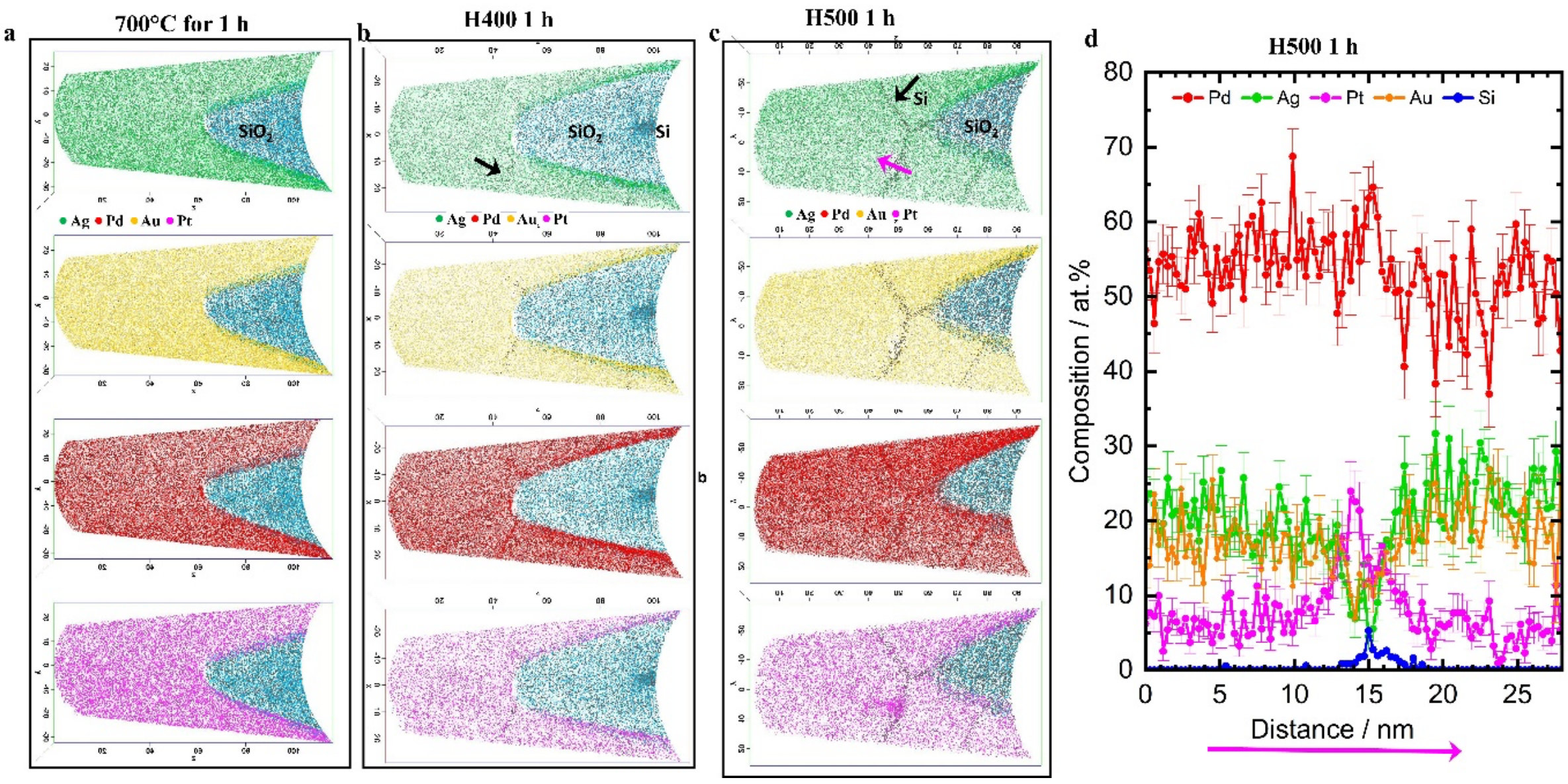

**Extended Data Fig. 9** Full elemental distributions of the homogenized and re-annealed states and Si segregation at GBs. **a-c**, APT reconstructions after homogenization at 700°C for 1 h (a) and subsequent annealing at 400°C (b, H400) and 500°C (c, H500) for 1 h, showing the full Ag, Au, Pd, Pt and Si distributions corresponding to the selected elemental maps presented in Fig. 3. Local Si enrichment reveals GBs in both H400 and H500. **d**, 1D composition profile across the GB indicated in c by the arrow, showing localized Si enrichment together with the Ag, Au, Pd and Pt distributions. The pink arrow indicates the profile direction, and the black arrows indicated the GBs with Si.

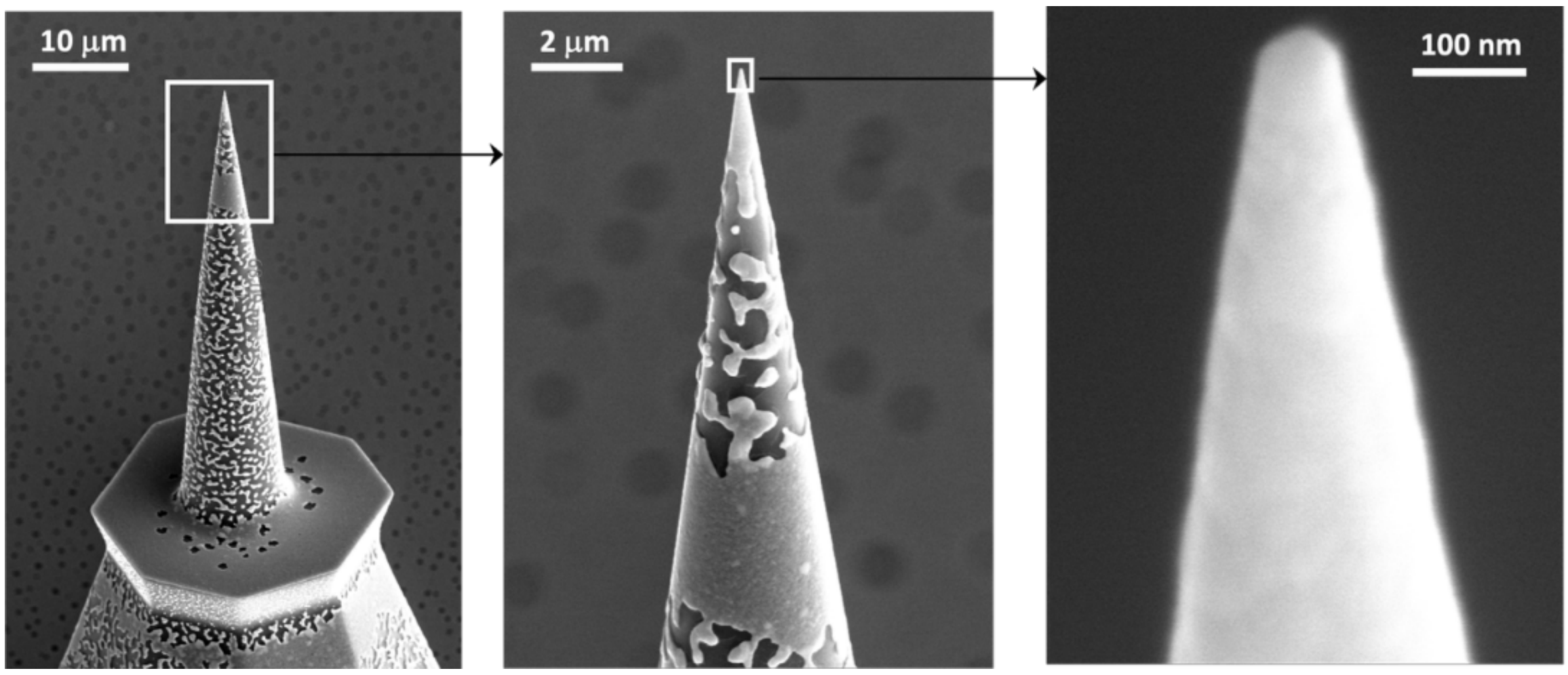

**Extended Data Fig. 10 Morphological stability through suppression of dewetting at the APT tip apex.** Scanning electron microscopy images at increasing magnification after annealing at 700°C for 1 h. The thin film dewets into isolated features along the surrounding shank, whereas the nanoscale alloy volume at the tip apex retains its continuous morphology.

**Extended Data Table 1** Molar volume and surface energy[52–54] of the current noble alloying elements.

| Element | Composition at% | Molar volume $V_m$ $10^{-6}$ $m^3/mol$ | Surface energy $\gamma$ $J/m^2$ |
|---|---|---|---|
| Ag | 23.6 | 10.3 | 1.5 |
| Au | 20.0 | 10.2 | 1.2 |
| Pd | 50.4 | 8.9 | 2.0 |
| Pt | 6.0 | 9.1 | 2.5 |

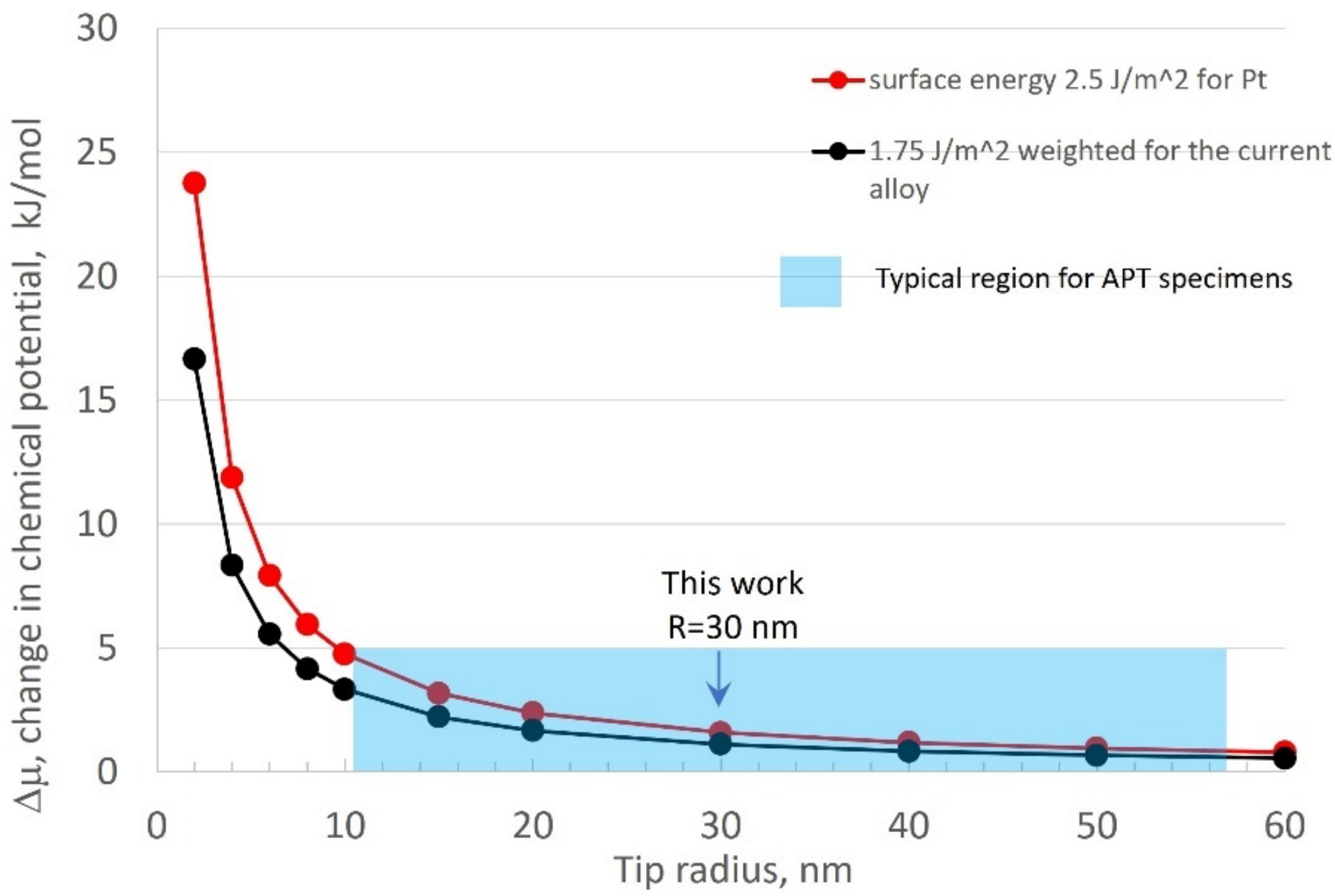

**Extended Data Fig. 11** Change in chemical potential due to Gibbs-Thomson effect.